\documentclass{rsos}

\usepackage[final]{pdfpages}

\begin{document}

\title{Charting the life of Billboard hits through memory, turnover, and predictability}

\author{
Arthur A. B. Pessa$^{1}$, 
Angelo A. Flores$^{1}$, 
Matja{\v z} Perc$^{2,3,4,5}$ 
and Haroldo V. Ribeiro$^{1}$ 
}

\address{$^{1}$Departamento de F\'isica, Universidade Estadual de Maring\'a -- Maring\'a, PR 87020-900, Brazil\\
$^{2}$Faculty of Natural Sciences and Mathematics, University of Maribor, Koro{\v s}ka cesta 160, 2000 Maribor, Slovenia\\
$^{3}$Community Healthcare Center Dr. Adolf Drolc Maribor, Ulica talcev 9, 2000 Maribor, Slovenia\\
$^{4}$Department of Physics, Kyung Hee University, Dongdaemun-gu, Seoul 02447, Republic of Korea\\
$^{5}$University College, Korea University, Seongbuk-gu, Seoul 02841, Republic of Korea
}

\subject{Complexity}
\corres{Arthur A. B. Pessa\\
\email{aabpessa@uem.br}}

\begin{abstract}
\small
Rankings shape the visibility and success of cultural products, yet their temporal dynamics remain underexplored when comparing distinct ranked objects within the same domain. Here, we use nearly seven decades of Billboard Hot 100 songs and six decades of Billboard 200 albums to investigate how success emerges, persists, and differs between songs and albums. We find that albums exhibit a heavier-tailed permanence distribution and reenter the charts more often than songs, whereas songs typically have longer uninterrupted runs. Similarity between successive charts decays much faster for songs than for albums, suggesting that individual hits reflect shorter-lived collective attention, while albums retain longer cultural memory. Rank-turbulence divergence shows that consecutive charts are similar, but that top positions are dominated more by rank reshuffling than by turnover. Entropy-based analyses reveal high uncertainty in rank movements, with distinct historical patterns for songs and albums and a strong dependence on trajectory length. Clustering of trajectories shows that chart success is organized into a small number of typical pathways, including canonical rise-and-fall trajectories, high-end persistence, and monotonic decline. Together, these results show that musical charts are not merely records of popularity, but dynamic memory systems in which attention, turnover, and predictability interact differently for songs and albums.
\end{abstract}

\maketitle

\section{Introduction}

Rankings of all sorts of objects are a natural way to establish hierarchies of popularity, influence, performance, and prestige~\cite{yucesoy2018sucees, clauset2015systematic, morgan2018prestige, erdi2020ranking}. As such, they provide relative assessments of relevance (or fitness), enable knowledge extraction from large collections of items, and frequently mediate access to collective attention and resources~\cite{erdi2020ranking, iniguez2022dynamics}. Furthermore, the underlying ratings or scores that generate rankings are often unavailable, making rank order the primary observable for comparative analyses~\cite{erdi2020ranking, dodds2023allotaxonometry}. At the same time, rankings can be self-reinforcing: high-ranked items tend to receive disproportionate visibility and demand, which in turn helps them remain highly ranked, introducing feedback loops between position and subsequent success~\cite{merton1968matthew, salganik2006experimental, perc2014matthew, joachims2017accurately}. 

Musical charts are a notable example of widely used rankings which have long served as a public proxy for popularity and commercial success of songs, albums, and artists~\cite{billboardHot100, billboard200, billboarArtist100}. Over the past decades, a growing body of literature has used chart-related data to study success in popular music, either through large-scale feature-based analyses or through the dynamics taking place inside the charts~\cite{mauch2015theevolution, interiano2018musical, shin2018on-chart, gourevitch2023billboard, bradlow2001abayesian, soh2024long-term, lech2025isitgetting, bhattacharjee2007stochastic, lao2016one-hit, davies2002theindividual, carroll2015notsolonely}. For instance, utilizing an evolutionary perspective, Mauch et al.~\cite{mauch2015theevolution} analyzed short excerpts of songs that reached the Billboard Hot 100 between 1960 and 2010, arguing that shifts in popular music are largely explained by changes in the relative popularity of styles, with limited evidence of homogenization or recurring cycles. Interiano et al.~\cite{interiano2018musical} operationalized success for more than 500{,}000 songs released between 1985 and 2015 as reaching the UK top charts and, combining acoustic features with statistical learning and a ``superstar'' indicator, predicted charting with high accuracy (${\sim}$80\%). Using South Korean charts, Shin and Park~\cite{shin2018on-chart} reported evidence that external factors, such as production companies and marketing campaigns, influence the success trajectories of K-Pop songs. Gour\'evitch~\cite{gourevitch2023billboard} investigated the success of artists in the Billboard 200 and found that their cumulative success follows a lognormal distribution. Several works have investigated the shapes of rank sequences within the Hot 100, either by explicitly modeling rank dynamics~\cite{bradlow2001abayesian, soh2024long-term} or by relating archetypal trajectories to success~\cite{lech2025isitgetting}. Finally, survival analyses~\cite{bhattacharjee2007stochastic, lao2016one-hit} showed changes in the survival probabilities of songs and albums on Billboard charts before and after technological shifts in the means of music distribution (LPs, cassette tapes, CDs, and so on). 

Taken together, these previous studies show that chart data provide a rich and nuanced picture of success and competition for attention. However, less is known about how songs and albums compare as ranked cultural products. Here, we present an analysis of the Billboard Hot 100 and Billboard 200, treating them as two long-running ranking systems associated with distinct but related cultural items. 
Beyond their long historical coverage and regular weekly publication, the analysis of these Billboard charts is especially interesting because of their role inside the music industry, working as a market-information source while also shaping commercial music~\cite{anand2000when}.
Our goal is to compare how songs and albums persist, reenter, and evolve within these charts, and to assess the extent to which their dynamics are shaped by randomization (chart reshuffling) and innovation (chart turnover). To this end, we first examine permanence distributions and the average decay of chart similarity. We then use rank-turbulence divergence~\cite{dodds2023allotaxonometry} to quantify week-to-week dissimilarity and to separate the contributions of chart reshuffling from those of turnover, both for the charts as a whole and emphasizing their top ranks. Next, we analyze the randomness of ensembles of rank trajectories grouped by length and year using an entropy-based approach~\cite{bandt2002permutation, bian2012modified, pessa2021ordpy, keller2003symbolic} and randomized baselines to assess predictability. Finally, we show that these trajectories can be grouped into four meaningful, archetypal, average shapes for each chart. 

The remainder of this work is organized as follows. First, we present our results regarding permanence, reentries, memory decay, rank divergence, entropy-based ensemble predictability, and trajectory shapes within the Billboard Hot 100 and Billboard 200 charts. We proceed by contextualizing and summarizing our findings, before outlining their implications. Our data are described at the end of the manuscript, along with details of the methods used to compare rankings and analyze rank trajectories.

\section{Results}\label{sec:results}

\subsection{Permanence and memory in musical rankings}

\begin{figure*}[!t]
\includegraphics[width=.95\linewidth]{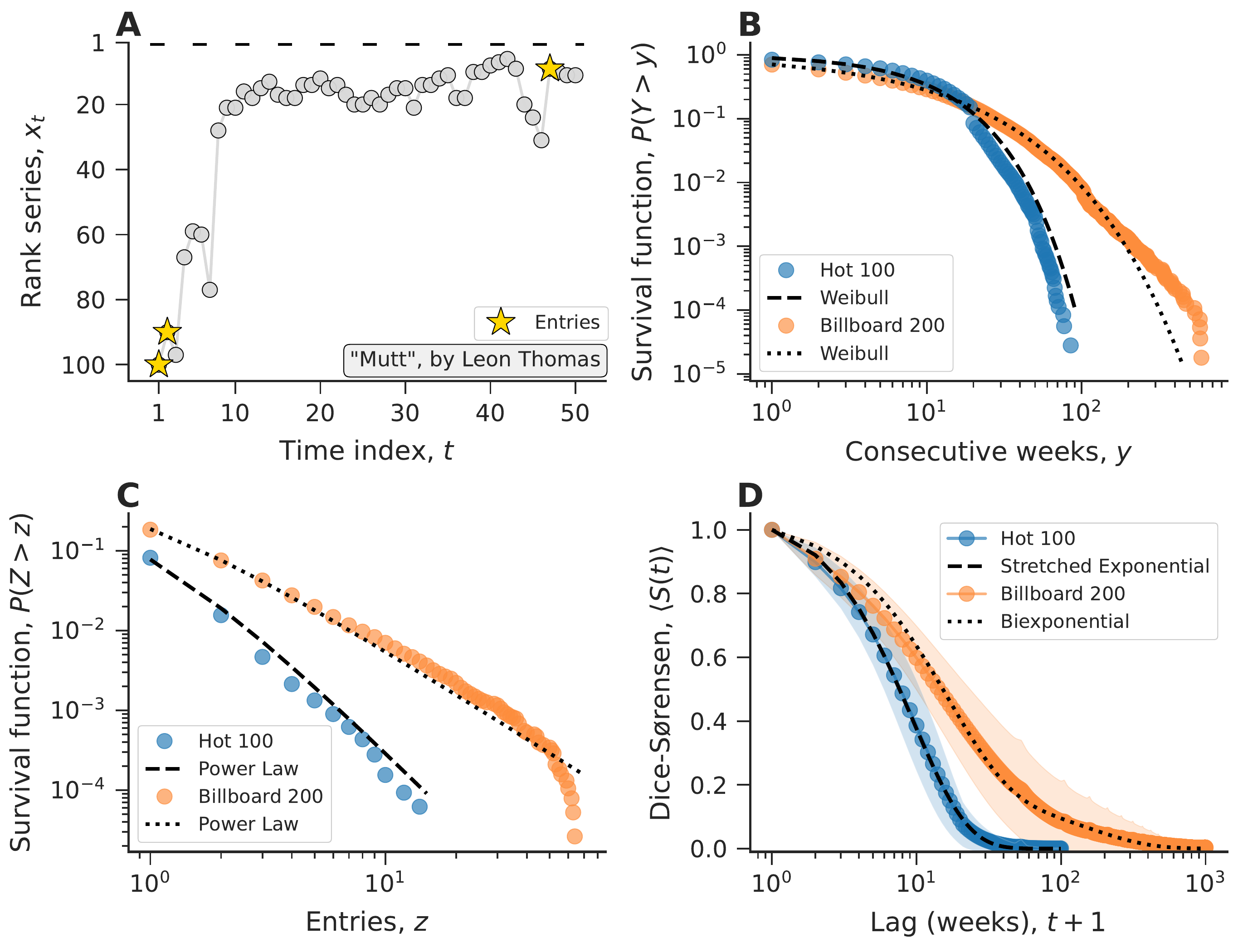}
\caption{Permanence and memory within Billboard charts. (A) Sequence of chart positions (in the Hot 100) for the song ``Mutt'', by Leon Thomas, in the weeks between 8 Feb 2025 and 31 Jan 2026. Every observation for which the song climbs into the top 100 songs is marked by a yellow star. (B) Complementary cumulative distributions of total consecutive weeks spent by songs and albums on the Billboard Hot 100 and Billboard 200 charts. Empirical distributions are shown by colored markers, while fits of discrete Weibull distributions are depicted by black dashed lines. (C) Empirical complementary cumulative distributions of the entries of songs and albums in the Billboard charts. Empirical distributions are shown by colored markers, while fits of discrete power-law distributions are depicted by black dashed lines. Notice that every entry of a song or album on a top ranking precedes a necessary exit, so distributions of entries/exits are equal. (D) Average Dice-S\o{}rensen coefficient for all pairs of songs (Billboard Hot 100) and albums (Billboard 200) rankings as a function of the increasing number of weeks (lag) between them. Empirical values are shown by colored markers, while black dashed lines present fits of stretched exponential and biexponential functions.}
\label{fig:1}
\end{figure*}

We start our analysis by investigating distributions of uninterrupted permanence of songs and albums on Billboard charts. Considering the total time span a song or album spends on the charts, we break this larger time interval into smaller intervals of contiguous weeks for which a song or album remained on a musical chart, thus accounting for possible exits and future reentries. Application of this criterion leads to a total of 35,803 rank series (or trajectories) for songs and 55,821 rank series for albums, greater numbers than those of unique songs or albums that have ever reached these charts (see Section~\ref{sec:data}). As an example, Figure~\ref{fig:1}{A} displays the chart positions of the song ``Mutt'', by Leon Thomas, the song with the longest stay (46 weeks) on the Hot 100 in the year 2025, and a total stay of 50 weeks between 8 Feb 2025 and 31 Jan 2026. We can see that the first time the song reached the top chart (100th position on 8 Feb 2025), it fell right off.  However, after it reached the Hot 100 again (90th position on 22 Feb 2025), the tune registered an uninterrupted permanence of 45 weeks, until it left the top 100 songs on 27 Dec 2025, just to reenter on 10 Jan 2026. By our approach, this total stay of ``Mutt'' in the Hot 100 is broken into three smaller rank series counting 1, 45, and 4 observations, respectively. The distributions of contiguous permanence for both Billboard charts are shown in Figure~\ref{fig:1}{B}. For typical values, the mean uninterrupted permanence for songs is 9.8 weeks and 11 weeks for albums. In addition, because these distributions are highly skewed, about 53\% of songs spend fewer than 8 weeks on the Hot 100, while the same fraction of albums remains in the Billboard 200 for fewer than 4 weeks. At the extremes, we find that the longest contiguous chart run in the Hot 100 is a 90-week series (Feb 2021 to Oct 2022) for ``Heat Waves'' by Glass Animals, whereas \textit{Legend: The Best Of Bob Marley And The Wailers} by Bob Marley \& the Wailers spent 631 consecutive weeks (Jan 2011 to Jan 2026) on the Billboard 200.

As we can further notice, the permanence of albums presents a longer tail, about one order of magnitude greater than that of songs. Despite this, the medians suggest that, if we randomly select an album and a song from these charts, the song might remain on the Hot 100 for a greater number of consecutive weeks than the album on the Billboard 200. Application of the Mann-Whitney test~\cite{conover1999practical} allows the rejection of the null hypothesis that albums stay for longer contiguous periods of time than songs on Billboard charts ($p = 0.0$, one-sided Mann-Whitney test). Given, however, that the Billboard 200 chart is twice as large as the Hot 100, we have built a ``Billboard 100'' chart (see electronic supplementary material, Figure~S1) with the permanence of the top 100 weekly albums to test the same hypothesis. Once again, the null hypothesis is rejected ($p = 0.0$, one-sided Mann-Whitney test).

Since we are examining contiguous time intervals spent by songs or albums on Billboard charts, it is also helpful to investigate how many times they enter and later exit these charts. Figure~\ref{fig:1}{C} shows the probabilities of songs and albums climbing into the top charts more than once. We see that songs are considerably more likely than albums to drop off the charts to never return again. Consistent with this, only about 8.2\% of songs have reentered the charts at least once, whereas more than twice as many albums (18.4\%) have done so. In this case, application of the Mann-Whitney test supports the null hypothesis of a greater number of reentries in the Billboard 200 (and ``Billboard 100'') than in the Hot 100 ($p = 1.0$, one-sided Mann-Whitney test). The upper tails also differ somewhat impressively, with ``Rockin' Around The Christmas Tree'' by singer Brenda Lee (and ``All I Want For Christmas Is You'' by Mariah Carey) entering and exiting the charts 15 times, whereas the album \textit{The Dark Side Of The Moon} by the band Pink Floyd has done the same in 70 occasions (since 1973). For songs, especially, seasonality appears to be a key driver of reentries~\cite{shin2018on-chart}, as a simple inspection of their titles reveals Christmas-related themes in 28 of the 43 songs with six or more chart entrances.

For the permanence and entries distributions, we also attempt to model their behavior using appropriate probabilistic models. Based on the shapes of these distributions and on the literature about heavy-tailed distributions in complex systems~\cite{clauset2009powerlaw, newman2005powerlaws}, we have tested whether these discrete distributions can be properly described by exponential, stretched exponential (Weibull), Yule, powerlaw, or (discretized) lognormal models. Interestingly, using the bootstrap test framework proposed by Clauset, Shalizi, and Newman~\cite{clauset2009powerlaw}, none of these models are found to be statistically acceptable descriptions of permanence or entries distributions (electronic supplementary material, Tables.~S1 and~S2). We believe this resistance to a description by single statistical models is partly due to changes in the lifespan of songs and albums within these charts, changes in the rules for songs or albums to reach (and stay on) the Billboard charts, and changes in the means of music consumption (factors widely reported in the literature~\cite{carroll2015notsolonely, lao2016one-hit, gourevitch2023billboard, lech2025isitgetting}), which renders our distributions amalgamations of several different regimes.

Despite these results, Figs.~\ref{fig:1}{B} and \ref{fig:1}{C} show the model distributions with the smallest AIC values and Kolmogorov-Smirnov distances to the actual data (electronic supplementary material, Tables~S1 and~S2). We notice that permanence in both the songs and albums charts is best described by discrete Weibull models $[P(X>x) = e^{-\lambda x^\beta}]$~\cite{nakagawa1975discrete}, similarly to previous results found for the success of bands in UK top charts~\cite{davies2002theindividual}. Interestingly, however, comparisons between the best model $[P(X>x) = e^{-\lambda x}; \lambda = 0.12, \beta = 0.97]$ and the empirical data indicate an approximately superexponential tail for consecutive weeks spent by songs on the Hot 100. On the other hand, albums display a subexponential permanence decay best modeled by $P(X>x)=e^{-\lambda x^\beta}, \text{with\ }\lambda = 0.35, \text{and\ } \beta=0.57$. For chart entries, a discrete power-law model~\cite{clauset2009powerlaw}, $P(X>x) = \zeta(\alpha, x+1)/\zeta(\alpha, 1)$, defined in terms of the Hurwitz zeta function $[\zeta(\alpha, x) = \sum_{n=0}^\infty (n + x)^{-\alpha}]$, gives the closest descriptions. We find $\alpha = 3.97$ for songs and $\alpha = 2.87$ for albums, compatible with the considerable differences in their tails. 

Beyond studying these overall, static distributions, our data also allow us to evaluate global temporal aspects such as evolution of the mean pairwise chart similarity. Figure~\ref{fig:1}{D} shows the values of the average Dice-S\o{}rensen coefficient $\langle S(t)\rangle$~\cite{legendre1998numerical, levy2025aguide}, a measure of set overlap, calculated for all pairs of rankings separated by a number $t$ of weeks (see Section~\ref{sec:methods} for more details). Values of $\langle S(t) \rangle \cong 1$ indicate that, on average, pairs of charts separated by $t$ weeks are comprised of the same songs or albums, while $\langle S(t) \rangle \cong 0$ indicates there is almost no overlap. Observing Fig.~\ref{fig:1}{D}, we first notice that it takes much longer for $\langle S(t)\rangle$ to reach zero in the case of albums (about $1000{\ \rm weeks}$) than in the case of songs (about $100{\ \rm weeks}$). More interestingly, despite not representing a direct measure of listener recall or cultural recognition, the decay of the average set overlap between all pairs of rankings might be interpreted as a proxy for the decay in the collective memory of songs or albums. Such a fall from memory depends both on the total time span a song or album spends on a Billboard chart and on the number of times they enter/exit them. Considering this collective memory framework, we fit three models of memory decay~\cite{higham2017fame, candia2019universal} to the average Dice-S\o{}rensen curve, namely, exponential, stretched exponential, and biexponential models (electronic supplementary material, Figure~S2 and Table~S3). These models are then evaluated with the Akaike Information Criterion (AIC), the coefficient of determination $(R^2)$ and the standard error of the regression (SER). For the Billboard Hot 100, a stretched exponential function, $\langle S(t) \rangle= e^{-\alpha t^\beta} (\alpha = 0.08, \beta = 1.12)$, is the best fit. This model readily yields characteristic times such as the half-decay time $\langle S(t_{1/2})\rangle = 1/2$ and the characteristic time of exponential decay $\tau$ such that $\langle S(\tau)\rangle = 1/e$, which we find to be $t_{1/2} \cong 7$ weeks and $\tau \cong 9$ weeks. For the Billboard 200, a biexponential model $\langle S(t) \rangle = \frac{1}{p-q+r} [(p-q)e^{-(p+r) t} + re^{-q t}\ (p = 0.053, q = 0.006, r = 0.010)$, performs the best. In this parametrization, the sum $p + r$ is associated to the rate of decay of the communicative memory (oral transmission) of albums and $q$ to the rate of decay of their cultural memory (physical recordings)~\cite{candia2019universal}. Most interestingly, this model allows us to estimate a ``crossover time'', $t_c = \frac{1}{p+r-q}\log{\left( \frac{(p+r)(p-q)}{rq}\right)}$, from which cultural memory becomes the dominant mechanism over communicative memory, which we find to be $\cong 68$ weeks.

\subsection{Dynamics of musical rankings}\label{sec:rankings}

\begin{figure*}[!t]
\includegraphics[width=1.\linewidth]{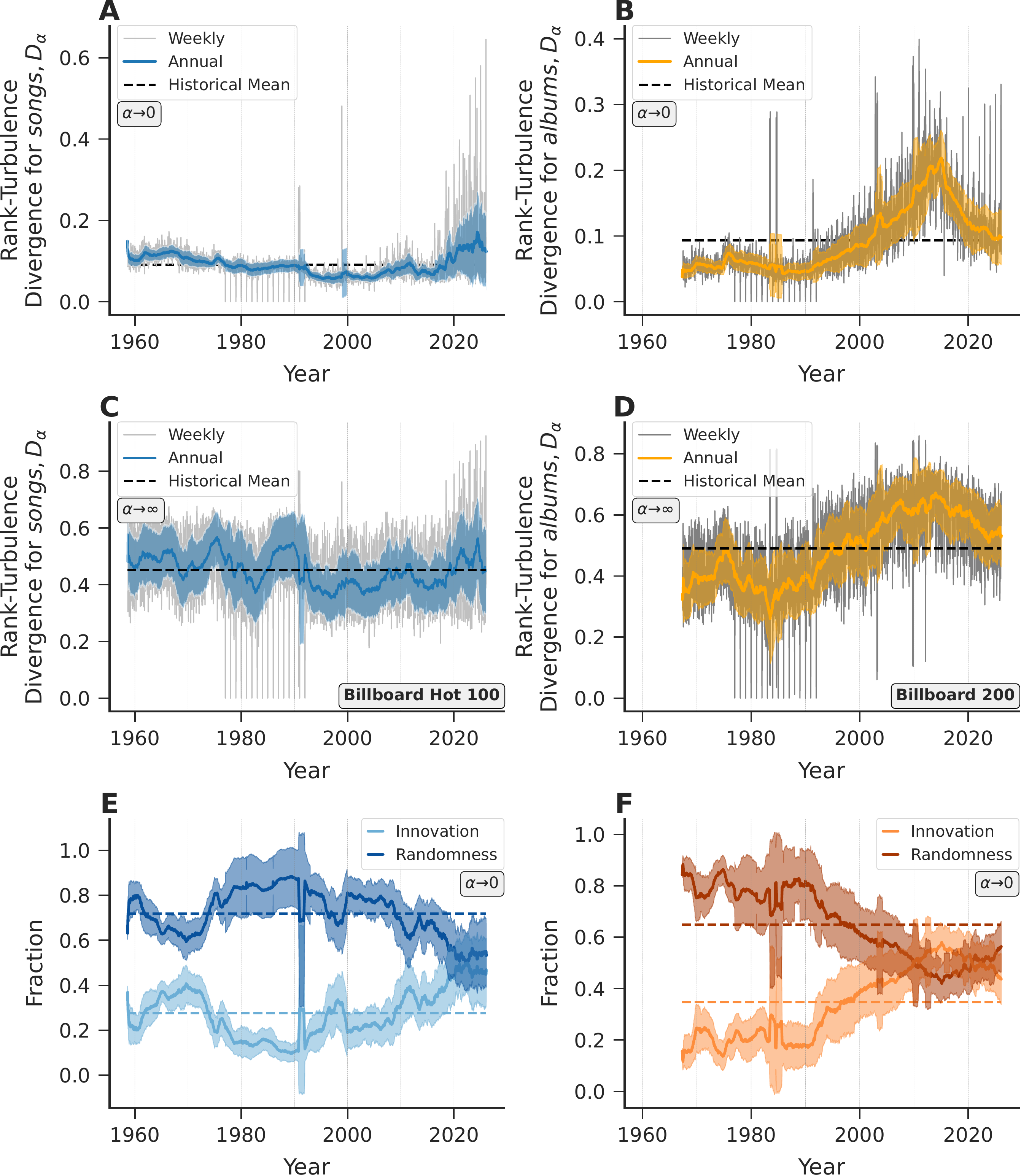}
\caption{Evolution of dissimilarity and mechanisms of rank change within Billboard charts. Rank-turbulence divergence when $\alpha \to 0$ for (A) songs and (B) albums in the Billboard Hot 100 and Billboard 200 charts. Rank-turbulence divergence when $\alpha \to \infty$ for (C) songs and (D) albums in Billboard rankings. In panels (A)-(D), weekly values for the rank-turbulence divergence are shown in gray, while 52-week (yearly) moving averages and standard deviations are colored. Historical averages are shown by black dashed lines. Randomness (reshuffling) and innovation (turnover) components of rank-turbulence divergence  $(\alpha \to 0)$ for (E) songs and (F) albums in the Billboard Hot 100 and Billboard 200 charts. In both panels (E) and (F), weekly values for these two components are omitted, and 52-week (yearly) moving averages and standard deviations are shown in colors. Historical averages are shown by dashed lines. In all panels, subtle vertical lines in the background indicate different decades traversed by the data.
}
\label{fig:2}
\end{figure*}

We continue our temporal analysis of the two Billboard charts by comparing temporally adjacent rankings. We do so by calculating the rank-turbulence divergence $(D_{\alpha})$~\cite{dodds2023allotaxonometry}, a versatile and recently proposed parametric statistic that allows for a detailed comparison of ranking pairs (see Section~\ref{sec:methods} for more details). Lower values of rank-divergence turbulence $(D_{\alpha} \cong 0)$ indicate identical rankings, while higher values $(D_{\alpha} \cong 1)$ indicate rankings formed by disjoint sets of items. In addition, the parameter $\alpha \in (0,\infty)$ creates the possibility of weighing differently the contributions of high- and low-ranked items to the overall rank-divergence statistic $(D_\alpha)$. Beyond the simple sharing of items (songs or albums) across time-adjacent charts, rank-divergence turbulence also captures the randomization (reshuffling of items shared by both rankings) and innovation (items climbing into or falling from the rankings) dynamics within the charts.

Figures~\ref{fig:2}{A} and~\ref{fig:2}{B} show that the values of $D_{\alpha}$ for week-adjacent rankings vary considerably. In these first results, we have also taken the limit $\alpha \to 0$ so that all items within the charts contribute equally to the rank-divergence turbulence statistic. The historical averages of rank-turbulence divergence for both charts are small and very close to each other ($\langle D_{0}\rangle = 0.09$ for both rankings), indicating that consecutive Billboard Hot 100 and Billboard 200 charts are typically very similar to each other. Furthermore, for songs, the yearly moving average of $D_0$ hovers closely around the historical mean, whereas for albums, the yearly moving average increases from the 1990s through the mid-2010s, then reverses and returns to the historical mean.

Figures~\ref{fig:2}{C} and~\ref{fig:2}{D} show the values of rank-turbulence divergence for all consecutive pairs of our two Billboard charts, but in the limit $\alpha \to \infty$, which allows us to investigate the dynamics at the top of the charts. In this limit, we find much higher historical averages, $\langle D_{\infty}\rangle = 0.45$ for the Hot 100 and $\langle D_{\infty}\rangle = 0.49$ for the Billboard 200. In addition, the yearly moving average of $D_{\infty}$ for the Hot 100 has oscillated around its historical mean, while yearly averages for the Billboard 200 have shown an ascending trend from the mid 1980s to the mid 2000s that has since reversed, leading to current values being close to the historical mean.

The differences between consecutive musical rankings can be further decomposed into contributions due to random changes $(D_{\alpha}^{\rm rand})$ in chart positions or due to the innovation $(D_{\alpha}^{\rm innov})$ of their falling from or climbing into the charts (see Section~\ref{sec:methods}). Figures~\ref{fig:2}{E} and~\ref{fig:2}{F} show the contributions of these two mechanisms, and we can see the Billboard Hot 100 was largely dominated, until 2010, by rank randomization of songs within the chart, but since then it has moved to a much more symmetric contribution, confirming an increase in turnover that has been noted in the literature~\cite{lech2025isitgetting}. On the other hand, periods of stability or trends in the rank-divergence turbulence of Billboard 200 (Fig.~\ref{fig:2}{A}) appear to be reflected in a diminishing randomization and increasing innovation in the album's chart that took place from 1990 to the mid 2010s, but that has since stabilized around an equal contribution of these two mechanisms. Repeating this same analysis, but prioritizing the top-ranked items $(\alpha \to \infty)$, we find much more stable contributions, with randomness dominating the dynamics at the top of both the songs and albums charts since their inceptions ($\langle D^{\rm innov}_{\infty}\rangle = 0.92$ for the Hot 100 and $\langle D^{\rm innov}_{\infty} \rangle = 0.86$ for the Billboard 200, see electronic supplementary material, Figure~S3).

\subsection{Randomness in sequences of rank positions}\label{sec:mpe}

\begin{figure*}[!t]
\includegraphics[width=1.\linewidth]{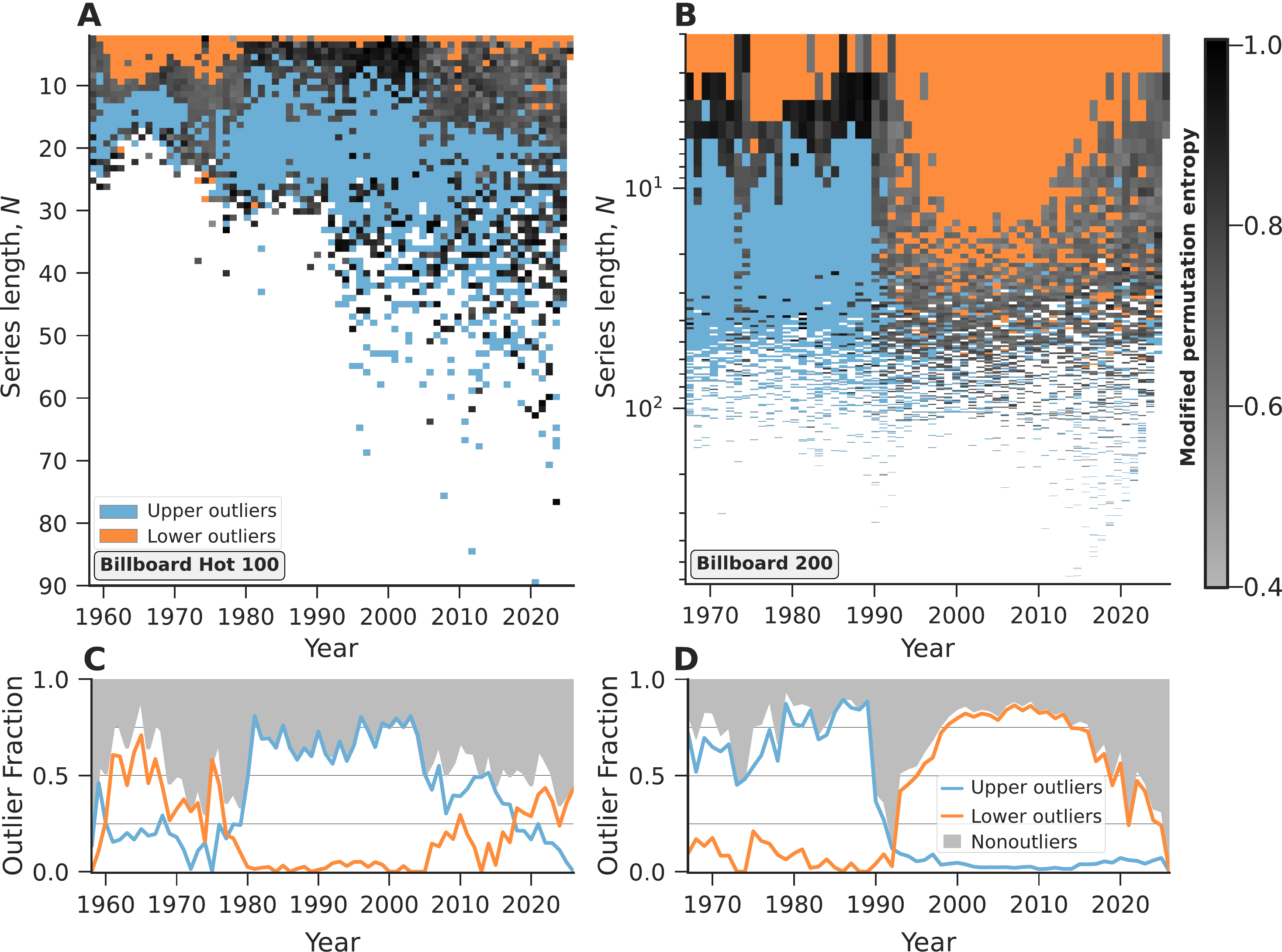}
\caption{Randomness of rank series within Billboard charts. Modified permutation entropy for length-year ensembles of rank series of (A) songs and (B) albums that reached the Billboard Hot 100 and Billboard 200 charts. Darker shades point to higher values of permutation entropy (randomness) and vice versa, as shown in the colorbar aside the panels. Logarithmic scales are used for the vertical axis of panels (B) due to the longer range of permanence of albums in the Billboard 200 chart. In both panels, ensembles are colored blue (orange) when deemed upper or lower outliers based on comparisons with randomized versions. Fractions of rank series of (C) songs and (D) albums pertaining to outlier ensembles per chart year (for all lengths). In both panels, gray shaded regions indicate the proportion of series belonging to ensembles that are not outliers.}
\label{fig:3}
\end{figure*}

After investigating the overall dynamics of the Billboard rankings, we next ask whether the predictability of rank trajectories has changed over time. Because most contiguous rank series are short, typically containing fewer than ten observations, their individual statistics might not be reliable. To circumvent this, for all series with at least two observations, we group them by length and by the year of their first observation. For each ensemble, we examine whether successive rank changes are distributed approximately equally among upward, downward, and sideways moves, which is effectively done by computing the group's modified permutation entropy, $H$~\cite{bian2012modified} (see Section~\ref{sec:methods} for details). When upward, downward, and sideways moves occur with nearly equal probabilities, $H \approx 1$; conversely, $H \approx 0$ indicates that one type of move dominates. Overall, we find high values of $H$ across groups of different sizes and years, with mean values $\langle H \rangle = 0.82$ for songs and $\langle H \rangle = 0.77$ for albums, indicating substantial uncertainty in rank movements.

To deepen our insights and provide a basis for comparison, we have estimated the average modified permutation entropy across 1,000 randomizations of the rank series within each length-year ensemble. Divergences from the average random values are then evaluated with Tukey's fences rule~\cite{tukey1977exploratory}, that is, in cases an ensemble's entropy lies above (below) the median plus (minus) 1.5 times the interquartile range of the random distribution of entropies obtained from its 1,000 randomizations, the ensemble behaves as an outlier to the expected randomness of the rank trajectories. Results of these comparisons are shown in Figure~\ref{fig:3}.

Figures~\ref{fig:3}{A} and~\ref{fig:3}{C} show a predominance of high-entropy ensembles for the Billboard Hot 100 between the 1980s and the mid-2000s. During this period, more than 50\% of all rank trajectories belonged to ensembles with higher entropy than their randomized counterparts. These characteristically high entropies are especially manifested for contiguous rank series longer than 9 observations, whereas shorter series are frequently outliers in the opposite direction (electronic supplementary material, Figure~S4). This seemingly counterintuitive result arises from our ternary encoding (up, down, sideways) of rank trajectories: randomization disrupts sideways moves (ties), effectively reducing their prevalence and shifting probability mass toward upward and downward changes, lowering the entropy of randomized ensembles. To confirm this, we overlook draws and consider them as equivalent to downward moves (see Section~\ref{sec:methods}). Performing such an analysis, most ensembles become lower (predictability) outliers (electronic supplementary material, Figure~S5). This way, we can conclude that sideways moves are an integral part of a song's chart life in a highly competitive attention market.

Figures~\ref{fig:3}{B} and~\ref{fig:3}{D} inform of somewhat different tendencies for the albums in the Billboard 200. From the chart inception until the late 1980s, most year-length ensembles are found to be upper (randomness) outliers, especially sequences longer than $10$ rank positions (electronic supplementary material, Figure~S4). Curiously, however, outliers of predictability became increasingly common, peaking just before the year 2000 and after 2010, with more than 75\% of all albums' rank sequences belonging to such ensembles. As with songs, shorter rank series of albums are frequently found in less-entropic ensembles. Also, sideways moves (ties) are a crucial part of the dynamics of their rank sequences (electronic supplementary material, Figure~S5).

\subsection{Typical shapes of rank sequences}\label{sec:shape}

In addition to revealing deviations of ensembles of rank series from their expected random behaviors, the study of the ups, downs, and lateral moves within Billboard charts also allows us to look for some typical shapes presented by these rank trajectories~\cite{bradlow2001abayesian, shin2018on-chart, soh2024long-term, lech2025isitgetting}. To do so, we concern ourselves with contiguous series counting at least three observations, because series with two observations are obliged to behave like straight lines with various inclinations. 

Since rank trajectories can be represented as sequences of three types of moves, we estimate the corresponding symbol probabilities via series symbolization. Positive rank changes (worsening rank) correspond to downward movement, here associated with the binary symbol 01; negative rank changes (improving rank) correspond to upward movement and the symbol 10; and rank permanence corresponds to 00 (see Section~\ref{sec:methods} for details). We then use the resulting three-dimensional feature vector of symbol probabilities as input to an agglomerative hierarchical clustering algorithm with Ward's linkage criterion, which iteratively merges clusters so as to minimize the increase in within-cluster variance~\cite{james2023introduction, mehta2019ahighbias}. To select the most adequate number of shape clusters $k$, we systematically increase $k$ from 1 to 7 and inspect their average trajectories (archetypal shapes). We retain a larger number of groups only when it yields additional, clearly distinct, and well-populated clusters. If increasing $k$ merely subdivides existing clusters into smaller groups with average shapes that are minor (or no) refinements of previously observed ones, we regard the extra complexity as uninformative. Under these precepts, we find that the Billboard Hot 100 exhibits meaningful changes in cluster-average shapes up to $k \leq 6$, whereas for the Billboard 200 such changes are limited to $k \leq 4$. For larger values of $k$, the algorithm primarily produces small, non-representative clusters whose average trajectories closely resemble those of existing groups (electronic supplementary material, Figures~S6 and~S7).

\begin{figure*}[!t]
\includegraphics[width=1\linewidth]{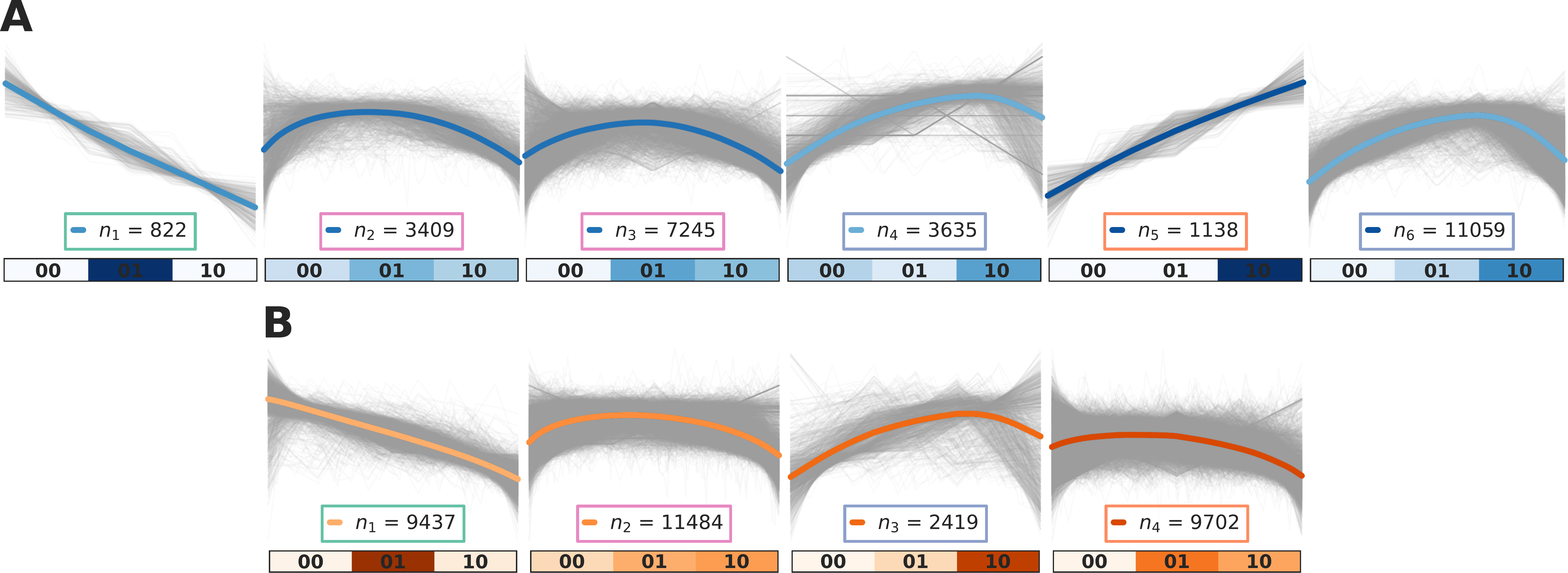}
\caption{Typical shapes of rank trajectories within Billboard charts. (A) Shapes of rank trajectories for the best grouping of songs' rank series into six clusters. Average shapes within clusters are colored, while individual trajectories are shown in gray. The six original clusters are further regrouped (based on their average shapes) into four archetypal shapes: descending (green), low-end (pink), high-end (dark blue), and ascending (orange). (B) Shapes of rank trajectories for the best grouping of albums' rank series into four clusters. Average shapes within clusters are colored, while individual trajectories are shown in gray. The four archetypal shapes of albums' trajectories are named descending (green), canonical (pink), high-end (dark blue), and low-end (orange). In all panels, individual rank trajectories are interpolated to render all series with the same length. The number of series $(n)$ belonging to each cluster is shown within the panels. For all clusters, the average probabilities of sideways (00), downward (01), and upward (10) moves is shown below the clusters, with darker (lighter) shades indicating greater (smaller) probabilities.}
\label{fig:4}
\end{figure*}

For the Hot 100, even at $k=6$, we can still merge two pairs of clusters, yielding only four representative archetypal shapes for rank trajectories. Figure~\ref{fig:4}{A} shows these four groups and their average shapes. We label them descending (cluster 1), canonical (clusters 2 and 3), high-end (clusters 4 and 6), and ascending (cluster 5). Notably, most song trajectories with more than two observations (27,308 series) fall into either the canonical cluster (10,654 series; 39\%) or the high-end cluster (14,694 series; 54\%). Both of these average shapes are closely aligned to the presumed canonical trajectory presented by songs' rank series, that is, a concave, ``inverted-U-like'' trajectory~\cite{shin2018on-chart, soh2024long-term, lech2025isitgetting}. These clusters, however, confer more information on the overall asymmetry of typical concave rank trajectories within the Hot 100. Furthermore, rank series assigned to different clusters are not uniformly distributed across lengths or years. The monotonically descending and ascending series are short, usually counting at most 10 observations (electronic supplementary material, Figure~S8). Ascending rank series were also much more common from chart inception (1958) to the late 1970s, while monotonically decreasing trajectories have become increasingly more common, especially since the year 2000. Otherwise, together, the two remaining clusters, which aggregate concave-like rank sequences, dominate the charts for all lengths and years.

For the Billboard 200, $k=4$ yields four distinct typical shapes representing albums' on-chart trajectories. Figure~\ref{fig:4}{B} highlights the average shapes of these clusters, which we name descending (cluster 1), canonical (cluster 2), high-end (cluster 3), and low-end (cluster 4). We can readily see that the majority of albums' rank series belong to the last three clusters (23,605 series; 71\%) whose average patterns are concave, canonical-like shapes~\cite{shin2018on-chart, soh2024long-term, lech2025isitgetting}. However, a large proportion of these rank sequences appear as monotonically decreasing curves (9,437 series; 29\%). Once again, trajectories belonging to different clusters have characteristic lengths and epochs. High-end shapes, beyond representing a small minority of series, are only found in rank sequences shorter than 30 observations, despite appearing throughout the whole history of the Billboard 200 chart (electronic supplementary material, Figure~S9). Conversely, the more interesting, monotonically decreasing series are found in trajectories with as many as 100 observations, and they are increasingly common in shorter series. In addition, these descending series were most common between the early 1990s and the early 2010s, reaching more than 50\% of all series in some years of the first decade of this century.

Finally, we notice that a more detailed look into some individual series within our final clusters on Figs.~\ref{fig:4}{A} and~\ref{fig:4}{B} points to member series whose shapes appear ``opposite'' to the average cluster shape. This phenomenon is a direct result of our clustering of probabilities for up, down, and sideways moves, since, for instance, a series with close-to-equal probabilities of these three moves might exhibit them in a ``U-like-shaped'' or ``inverse-U-like-shaped'' trajectory. A closer inspection of these member series, however, shows they are few within their clusters.

\section{Conclusions}\label{sec:conclusions}

Rankings can be understood as dimensionality-reduction tools that guide assessments of relative standings among classes of objects~\cite{erdi2020ranking, dodds2023allotaxonometry}. In the case of musical charts, this perspective is especially useful because chart position compresses multiple dimensions of cultural success into a single ordered list. Such success is known to depend not only on intrinsic attributes or perceived ``quality'', but also on social influence, language, marketing campaigns, and previous success~\cite{salganik2006experimental, shin2018on-chart, interiano2018musical, amaral2026breaking}. From this standpoint, we have presented a comparative and systematic analysis of the historical and temporal dynamics of two long-running musical rankings: the Billboard Hot 100, representing songs, and the Billboard 200, representing albums. 

Our results show that songs and albums follow markedly different regimes of chart persistence and collective memory. By investigating contiguous, uninterrupted chart permanence, we have found that albums display a much more right-skewed distribution than songs, with substantially longer extreme runs. Because this analysis was based on continuous intervals of presence in the charts, we also examined reentries and found that they are considerably more common for albums, again with a much heavier-tailed distribution. At the same time, permanence and reentry distributions resist description by a single historical curve, probably because they combine several regimes shaped by changing consumption technologies, chart rules, and commercial practices~\cite{bhattacharjee2007stochastic, lao2016one-hit, carroll2015notsolonely}. These differences point to distinct commercial and memory dynamics for songs and albums as cultural products~\cite{higham2017fame, candia2019universal}.

To make it possible to investigate memory dynamics of musical products through these ranks, we made an operational interpretation of the decreasing similarity of charts separated in time as a decay of collective memory and attention.
We found that the Billboard Hot 100 exhibits a fast, superexponential decrease, characterized by a half-decay time ($t_{1/2} \cong 7$ weeks) and an exponential decay time ($\tau \cong 9$ weeks). By contrast, the Billboard 200 is better described by a biexponential model, indicating that cultural memory becomes the dominant mechanism of album remembrance over oral, or communicative, memory after about 68 weeks on chart. 
In this sense, songs behave more like short-lived attention objects, whereas albums more clearly retain the signature of longer-term cultural memory. 
Using the rank-divergence turbulence statistic~\cite{dodds2023allotaxonometry}, we further showed that consecutive rankings are usually highly similar, but that the overall chart dynamics and the dynamics at the very top can be separated. Although both Billboard charts are now driven almost equally by rank reshuffling (randomness) and chart turnover (innovation), their top ranks have historically been dominated by the most successful items exchanging positions among themselves.

The analysis of upward, downward, and sideways movements within rank trajectories further revealed that sideways moves are not a marginal detail, but a crucial component of chart dynamics. Overall, randomness in year-length ensembles of continuous rank sequences appears to be more dependent on trajectory length for songs in the Billboard Hot 100, and more dependent on historical time for albums in the Billboard 200. Moreover, by decomposing contiguous rank series into these three types of rank movement, a clusterization approach based on move frequency allowed us to classify even very short trajectories, with a minimum of 3 observations. This analysis extends the usual view that musical chart trajectories are necessarily concave, or ``inverse-U-like''~\cite{shin2018on-chart, soh2024long-term, lech2025isitgetting}. Instead, chart success is organized into a small number of typical pathways, including canonical rise-and-fall, high-end persistence, and monotonically decreasing or increasing trajectories.

The distributions of cluster membership across trajectory length and historical year are far from uniform. Monotonic series are mostly short, which is consistent with the finding that ensembles of shorter trajectories are frequently outliers of predictability. For the Billboard 200, many rank-sequence ensembles were outliers in predictability between the 1990s and 2010s, and the overrepresentation of rank series from this period in the descending cluster is consistent with this broader temporal pattern. Together, these results suggest that musical charts are not merely records of popularity. They are dynamic systems of cultural attention in which persistence, turnover, memory, and predictability interact differently for songs and albums.

Future investigations may extend our framework to genre-specific charts or top charts from other countries. Another interesting direction of inquire would be to examine more directly how chart construction, platform design, and shifts in modes of music consumption reshape the balance among memory, turnover, and predictability in musical success.
Moreover, we believe our analyses could also be applied to other rankings of cultural products such as books and films or to rankings derived from streaming platforms or social media.

\section{Data}\label{sec:data}

Our data comes from the Billboard Hot 100 (songs) and the Billboard 200 (albums) weekly charts~\cite{billboardHot100, billboard200, billboard2026github}. Our Billboard Hot 100 charts span the period from 4 Aug 1958 to 31 Jan 2026 (3,522 weeks) and rank the 100 most popular songs in the United States. The Billboard 200 rankings extend from 15 Apr 1967 to 31 Jan 2026 (3,070 weeks) and list the 200 most popular albums in the United States. Together, these weekly rankings produce sets of 32,295 unique songs and 38,056 unique albums that appear at least once in these charts. Since songs and albums might enter (and exit) the charts more than once, we extract time series of rank positions (rank series) sampled at a weekly rate by considering contiguous periods of time they remain on Billboard charts. This approach leads to a total of 35,803 rank series (or trajectories) for songs and 55,821 rank series for albums, which are the time series we investigate throughout this work. All code and data necessary to reproduce the results and figures presented here are available at the GitLab repository \textbf{gitlab.com/arthurpessa/billboard-charts}.

\section{Methods}\label{sec:methods}

\subsection{Ranking-comparison statistics}\label{subsec:rank_stats}

We start by considering a pair of rankings (charts) $R_i$ and $R_j$ of the same type of item (either songs or albums). Naturally, we can think of rankings as lists such that, for instance, $R_i = (r_{1, i}, r_{2, i}, ..., r_{\tau, i}, r_{n_i, i})$, with $r_{\tau, i}$ the position of the $\tau$-th item in ranking $R_i$ and $n_i$ the total number of items constituting the ranking. For ease of notation, we have referenced an item within a ranking by its position, but notice that in our data an item is defined by the combination of artist name and song name (for the Hot 100) or artist name and album name (for the Billboard 200). Reinterpreting these previous rankings as sets, written as $\mathcal{R}_i$ and $\mathcal{R}_j$, their similarity (overlap) can be assessed by the Dice-S\o{}rensen coefficient $(S)$~\cite{legendre1998numerical, levy2025aguide}, defined as
\begin{equation}
    S(\mathcal{R}_i, \mathcal{R}_j) = \dfrac{2 |\mathcal{R}_{i,j}|}{|\mathcal{R}_i| + |\mathcal{R}_j|}\,,
\end{equation}
with $\mathcal{R}_{i,j} = \mathcal{R}_{i} \bigcup \mathcal{R}_{j}$ the union of rankings sets and the notation $|\cdot|$ indicating set cardinality. By definition, $S \in [0,1]$, with $S = 0$ indicating perfectly dissimilar sets (no overlap) and $S = 1$ perfectly similar (identical) sets~\cite{levy2025aguide}. In addition, to assess the average decay in chart similarity, we calculate $\langle S(t) \rangle$, the average value of the Dice-S\o{}rensen coefficient for all pairs of charts separated by $t$ weeks. 

To extend our analysis of ranking similarities and further separate the influences of rank reshuffling/randomization (items changing positions within rankings) and turnover/innovation (items entering or exiting rankings) within musical charts, we have utilized the rank-turbulence divergence (RTD) statistic~\cite{dodds2023allotaxonometry}. For a pair of rankings $R_i$ and $R_j$, rank-divergence turbulence $(D_{\alpha})$ is defined by
\begin{equation}\label{eq:rtd}
    D_{\alpha}(R_i\, ||\, R_j) = \sum_{\tau \in \mathcal{R}_{i,j}} \frac{1}{N_{i,j; \alpha}} \frac{\alpha + 1}{\alpha} \left| \frac{1}{[r_{\tau, i}]^\alpha} - \frac{1}{[r_{\tau, j}]^\alpha}\right|^{1/(\alpha + 1)} \,,
\end{equation}
where, once again, $\mathcal{R}_{i,j} = \mathcal{R}_{i} \bigcup \mathcal{R}_{j}$ and $r_{\tau, i}\, (r_{\tau, j})$ is the position of item $\tau$ within ranking $R_i\, (R_j)$. RTD is a parametric statistic that depends on the tunable parameter $\alpha \in (0, \infty)$ that further fixes the normalization constant $N_{i,j; \alpha}$ so that $D_{\alpha} \in [0,1]$. The lower limit of rank-divergence turbulence $(D_{\alpha} = 0)$ indicates a pair of identical rankings $(R_i = R_j)$, while its upper limit $(D_{\alpha} = 1)$ indicates a pair of rankings $R_i$ and $R_j$ formed by two completely different sets of items $(\mathcal{R}_i \bigcap \mathcal{R}_j = \varnothing)$. 

RTD deals with unshared items in the following way. If an item $\gamma$ pertains only to $R_i$ $(\gamma \notin \mathcal{R}_j)$, $\gamma$ is appended to the end of the $R_j$ ranking such that $r_{\gamma, j} = |\mathcal{R}_j| + 1$. In case several items from $R_i$ are not shared by $R_j$, these are appended, once again, at the end of $R_j$ and assigned the same average ranking $\langle r\rangle  = |\mathcal{R}_j| + \frac{1}{2} (1 + |\mathcal{R}_i - \mathcal{R}_j|)$, with $\mathcal{R}_i - \mathcal{R}_j$ the set of items inside $\mathcal{R}_i$ but outside $\mathcal{R}_j$.

The tuning parameter $\alpha$ plays the important role of weighing contributions of rank changes from different items based on their positions within the rankings. In case $\alpha \to 0$, rank changes in lower (larger $r_{\tau}$) and higher (smaller $r_{\tau}$) position items contribute equally to $D_{0}(R_i\, ||\, R_j)$. In this limit, the rank-turbulence divergence statistic ($D_{0}$) assumes the form
\begin{equation}
    D_{0}(R_i\, ||\, R_j) =  \sum_{\tau \in \mathcal{R}_{i,j}} \frac{1}{N_{i,j; \alpha \to 0}} \left| \ln{\frac{r_{\tau, i}}{r_{\tau, j}}} \right| 
\end{equation}
with the normalization constant
\begin{equation}
\begin{split}
    N_{i,j; \alpha \to 0} = &\sum_{\tau \in \mathcal{R}_i} \left| \ln{\frac{r_{\tau, i}}{|\mathcal{R}_i| + \frac{1}{2}(|\mathcal{R}_j|+1)}} \right| \\
    &+  \sum_{\tau \in \mathcal{R}_j} \left| \ln{\frac{r_{\tau, j}}{\frac{1}{2}(|\mathcal{R}_i|+1) + |\mathcal{R}_j|}} \right|\,.
\end{split}
\end{equation}
On the other hand, when $\alpha \to \infty$, rank changes in top-ranked items (smaller $r_{\tau}$) are the major contributors to the $D_{\infty}(R_i\, ||\, R_j)$ statistic. In this limit, the RTD statistic ($D_{\infty}$) can be written as
\begin{equation}
\begin{split}
    D_{\infty}(R_i\, ||\, R_j) = \sum_{\tau \in \mathcal{R}_{i,j}} \frac{1 - \delta_{r_{\tau, i} r_{\tau, j}}}{N_{i,j; \alpha \to \infty}}\,\max_{\tau}{\left( \frac{1}{r_{\tau, i}}, \frac{1}{r_{\tau, j}}\right)}\,,
\end{split}
\end{equation}
with the appropriate normalization constant $(N_{i,j; \alpha \to \infty})$ given by
\begin{equation}
    N_{i,j; \alpha \to \infty} = \sum_{\tau \in \mathcal{R}_i} \frac{1}{r_{\tau, i}} +  \sum_{\tau \in \mathcal{R}_j} \frac{1}{r_{\tau, j}}\,.
\end{equation}

Because $D_{\alpha}(R_i\,||\,R_j)$ is additive over items in the rankings, we might conveniently rewrite Eq.~\eqref{eq:rtd} as a sum of itemwise contributions~\cite{dodds2023allotaxonometry},
\begin{equation}
    D_{\alpha}(R_i\,||\,R_j)=\sum_{\tau\in\mathcal{R}_{i,j}} \delta D_{\tau,\alpha},
\end{equation}
where $\delta D_{\tau,\alpha}$ denotes the normalized contribution of item $\tau$ [as implicitly defined by the summand in Eq.~\eqref{eq:rtd}]. This decomposition enables us to separate the overall divergence into two parcels that reflect distinct mechanisms of chart change: \textit{(i)} rank randomization, corresponding to the reordering of items that remain present in both rankings, and \textit{(ii)} innovation/turnover, corresponding to items that are present in only one of the rankings due to their climbing into the charts or falling from them.

To define these contributions, given a pair of rankings $R_i$ and $R_j$, we first define the shared $(\mathcal{S}_{i,j})$ and exclusive $(\mathcal{E}_{i,j})$ sets as
\begin{equation}
\mathcal{S}_{i,j}=\mathcal{R}_i\cap\mathcal{R}_j
\end{equation}
and
\begin{equation}
\mathcal{E}_{i,j}=(\mathcal{R}_i - \mathcal{R}_j)\cup(\mathcal{R}_j - \mathcal{R}_i)\,.    
\end{equation}
We can now write the rank-turbulence divergence as a two-part contribution,
\begin{equation}
    D_{\alpha}(R_i\,||\,R_j)=D_{\alpha}^{\rm{rand}}(R_i\,||\,R_j)+D_{\alpha}^{\rm{innov}}(R_i\,||\,R_j)\,,
\end{equation}
where the randomization component is defined by
\begin{equation}
    D_{\alpha}^{\mathrm{rand}}(R_i\,||\,R_j)=\sum_{\tau\in\mathcal{S}_{i,j}} \delta D_{\tau,\alpha},
\end{equation}
and the innovation (turnover) component is
\begin{equation}
    D_{\alpha}^{\mathrm{innov}}(R_i\,||\,R_j)=\sum_{\tau\in\mathcal{E}_{i,j}} \delta D_{\tau,\alpha}\,.
\end{equation}

\subsection{Time series statistics}\label{subsec:ts_stats}

To study the predictability of sequences of rank positions of songs or albums within musical charts, we have utilized the modified permutation entropy (MPE)~\cite{bian2012modified}, a time series statistic belonging to the set of ordinal methods~\cite{pessa2021ordpy} originated from the symbolization approach put forward by Bandt and Pompe~\cite{bandt2002permutation}. 

To estimate MPE, we start by grouping our contiguous rank series by length and year of first observation. These year-length ensembles are then treated as $m$-dimensional time series $\{\mathbf{x}_t\}_{t = 1, ..., N}$, with $m$ the number of sequences within each group. Individual rank sequences belonging to the ensemble are given by $\{x^i_t\}_{t = 1, ..., N}$, with $i = 1, 2, ..., m$. The first step in the evaluation of modified permutation entropy consists of dividing each individual series $\{x^i_t\}$ into overlapping partitions $w_p^i$ defined by
\begin{equation}\label{eq:1dpartition}
    w^i_p =  (x^i_{p}, x^i_{p + 1}, x^i_{p + 2}, \dots, x^i_{p + (d - 2)}, x^i_{p + (d - 1)})\,,
\end{equation}
where $p = 1, \dots, N - (d - 1)$ and the parameter $d$, called embedding dimension~\cite{bandt2002permutation}, defines the length of these superposing windows. The choice of $d$ is constrained by the typical length of the series under investigation so that longer series might be studied with longer partitions~\cite{bandt2002permutation, bian2012modified, pessa2021ordpy}. Most commonly, $d \in \{2, 3, ..., 7\}$, but our very short rank series (typically $N < 10$) basically restricts us to setting $d = 2$.

After each rank sequence is partitioned with $d = 2$, each window $[w^i_p =  (x^i_{p}, x^i_{p + 1})]$ is mapped into a binary symbol depending on the relative amplitude of its rank positions. Considering, for instance, the first series of an ensemble, and its first partition, $w_1^1 = (x_1^1, x_2^1)$, we might have $x_1^1 = x_2^1$, $x_1^1 < x_2^1$, or $x_1^1 > x_2^1$. For each of these cases, we map the partition into one of three possible symbols: $\Pi_1 = 00$ (in case $x_1^1 = x_2^1$); $\Pi_2 = 01$ (if $x_1^1 < x_2^1$); $\Pi_3 = 10$ ($x_1^1 > x_2^1$). The mapping of all partitions into binary symbols leads to an $m$-dimensional symbolic sequence $\{\boldsymbol{\pi}_p\}_{p = 1, \dots, n}$ where each $\{\pi_p^i\}$ stands for the sequence of binary symbols associated with the $\{x^i_t\}$ series. Using these sequences, we estimate the probability $p_j^i$ of occurrence of a binary symbol $\Pi_j$ in the $i$-th series by calculating its relative frequency through
\begin{equation}\label{eq:permutation_probability}
    p^i_j(\Pi_j) = \frac{\text{number of partitions of type} \ \Pi_j \ \text{in} \ 
    \{\pi^i_p\}}{N - (d-1)}\,,
\end{equation}
for $j = 1,2,3$ and $\Pi_1 = 00$, $\Pi_2 = 01$, and $\Pi_3 = 10$. This way, we obtain probability distributions of binary symbols $P^i = \{p^i_j(\Pi_j)\}_{j = 1, 2, 3}$ for each time series $\{x^i_t\}_{t = 1,...,N}$ within an ensemble. Using these individual distributions, we can further combine them into a unique distribution for the multivariate series $\{\mathbf{x}_t\}_{t = 1, ..., N}$, a distribution of pooled symbols $P = \{p_j(\Pi_j)\}_{j = 1, 2, 3}$~\cite{keller2003symbolic} by doing
\begin{equation}\label{eq:pooled_probability}
    p_j(\Pi_j) = \frac{1}{m}\sum_{i = 1}^m p^i_j(\Pi_j)\,.
\end{equation}
Finally, using the discrete probability distribution of binary patterns $\{p_j(\Pi_j)\}_{j = 1,2,3}$, we estimate the (normalized) modified permutation entropy $(H)$~\cite{bian2012modified} defined as
\begin{equation}
    H = -\dfrac{1}{\ln{3}} \sum_{j = 1}^{3} p_j(\Pi_j) \ln{p_j(\Pi_j)}\,.
\end{equation}
The values of the modified permutation entropy are such that $H \in [0,1]$, with $H \cong 0$ indicating monotonic ensembles of rank series (always up, always down, or always sideways) and $H \cong 1$ pointing to ensembles with approximately equiprobable changes in rank moves.

To highlight the importance of sideways moves for rank series, we also estimate the permutation entropy~\cite{bandt2002permutation, keller2003symbolic} of the year-length ensembles. The crucial difference between permutation entropy and modified permutation entropy lies in the manner they treat draws (sideways moves in rankings). Following the literature~\cite{pessa2021ordpy}, when symbolizing partitions with observations of equal values $[w = (x^*, x^*)]$, they are symbolized as if the first observation is smaller than the second one, thus attributing the symbol $\Pi_2 = 01$ to these partitions. This way, permutation entropy is estimated only from the probabilities of the symbols $\Pi_2$ and $\Pi_3$ (downward and upward rank moves).

\section*{Ethics} 
This work did not require ethical approval from a human subject or animal welfare committee.

\section*{Data accessibility}
Code and data for replicating this study can be found at \url{gitlab.com/arthurpessa/billboard-charts}.

\section*{Declaration of AI use}
We have not used AI-assisted technologies in creating this article.

\section*{Authors' contributions}
A.A.B.P.: conceptualization, data curation, formal analysis, investigation, methodology, software, validation, visualization, writing-original draft, writing-review and editing. A.A.F.: conceptualization, data curation, formal analysis, investigation, methodology, software, validation, visualization, writing-original draft, writing-review and editing. M.P.: conceptualization, data curation, formal analysis, investigation, methodology, software, validation, visualization, writing-original draft, writing-review and editing. H.V.R: conceptualization, data curation, formal analysis, investigation, methodology, software, validation, visualization, writing-original draft, writing-review and editing.

All authors gave final approval for publication and agreed to be held accountable for the work performed therein.

\section*{Conflict of interest declaration} 
We declare we have no competing interests.

\section*{Funding}
We acknowledge the support of the Coordena\c{c}\~ao de Aperfei\c{c}oamento de Pessoal de N\'ivel Superior, the Conselho Nacional de Desenvolvimento Cient\'ifico e Tecnol\'ogico (CNPq -- Grants 303533/2021-8 and 317522/2025-6), and Slovenian Research and Innovation Agency (Javna agencija za znanstvenoraziskovalno in inovacijsko dejavnost Republike Slovenije) (Grant P1-0403).

\bibliographystyle{vancouver}
\bibliography{references}

\clearpage

\includepdf[pages=1-12,pagecommand={\thispagestyle{empty}}]{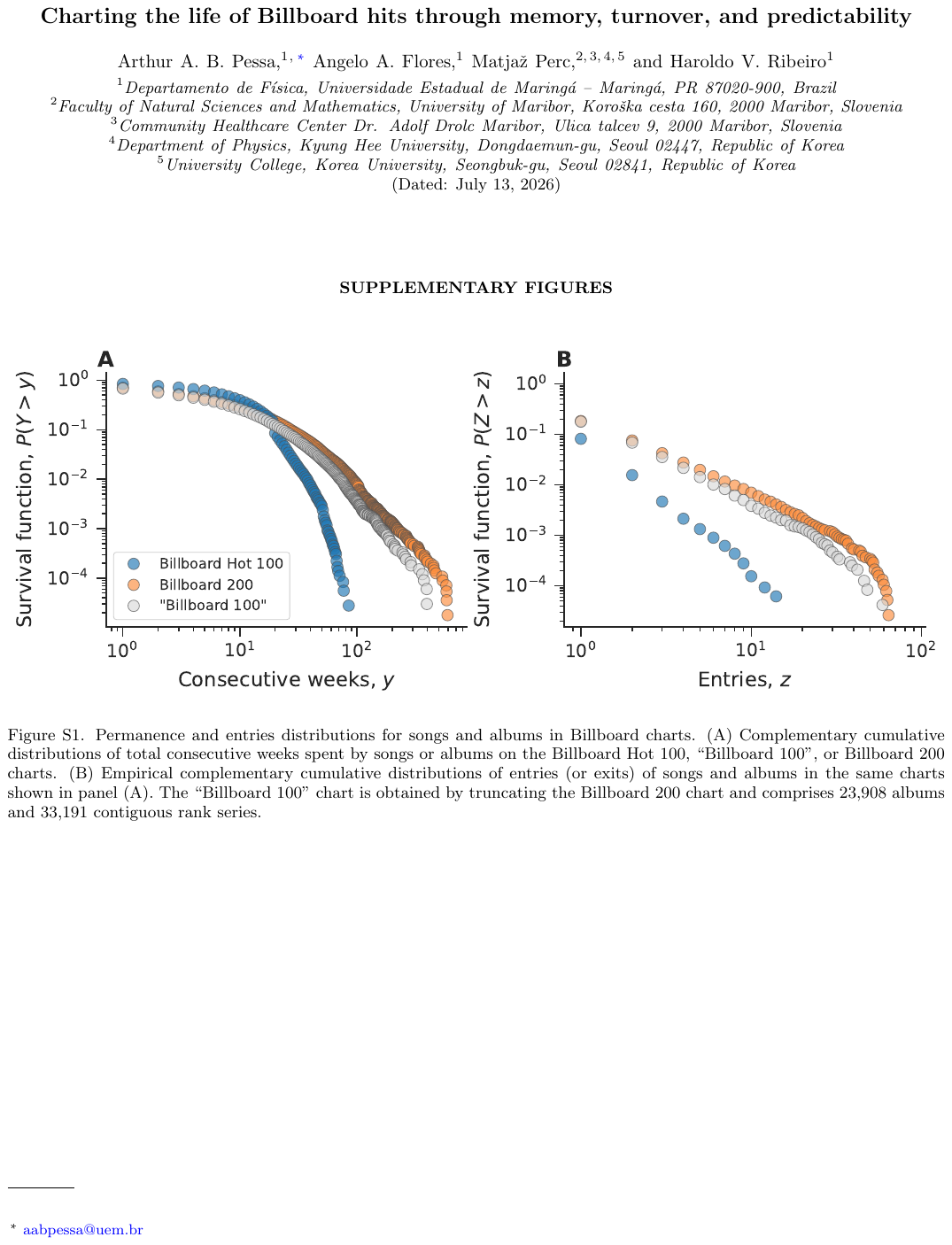}

\end{document}